\documentstyle[12pt,A4]{article}

\title{{\bf  RPA calculation of $K^+$-nucleus cross-sections
 with a density-dependent Brueckner-Hartree-Fock
NN interaction }}
\author{ J.C. CAILLON and J. LABARSOUQUE\\
Centre d'Etudes Nucl\'eaires de Bordeaux-Gradignan$^{\dag}$\\
 Universit\'e Bordeaux I\\ rue du
Solarium, 33175 Gradignan Cedex, France}
\date{}
\begin{document}
\begin{titlepage}
\maketitle
\thispagestyle{empty}
\begin{abstract}
In the calculation of the $K^+$-nucleus cross sections, the coupling
of the mesons exchanged between the $K^+$ and the target nucleons
to the polarization of the Fermi and Dirac seas has been
taken into account. This polarization has been calculated in the
one-loop approximation but summed up to all orders (RPA approximation).
For this calculation a density-dependent  Brueckner-Hartree-Fock NN
interaction providing a good description of both nuclear matter and finite
nuclei has been used.
The agreement with experiment is considerably improved.
\end{abstract}

PACS numbers: 25.80.Nv, 21.30.+y, 21.60.Jz, 24.10.Jv

\noindent E-mail LABARS@FRCPN11.IN2P3.FR \\
\footnotesize{$\dag$ UMR CNRS 6426}
\end{titlepage}
 \newpage
 These last years, significant progress have been made towards a better
understanding of the properties of both nuclear matter and finite nuclei
starting from a free-space nucleon-nucleon interaction. Using a relativistic
Brueckner-Hartree-Fock theory with a one-boson-exchange potential where
coupling constants and form factors were obtained from the NN scattering
data, Brockmann and Machleidt\cite{bm} reproduced the saturation properties
of nuclear matter nearly quantitatively. The scalar and vector fields they
obtained have then been introduced as input in a relativistic
 density-dependent
Hartree approach for finite nuclei by Brockmann and Toki\cite{bt} who obtained
a good agreement with experiment for both the binding energy and the
mean-square radius in $^{16}O$ and $^{40}Ca$. This result solves an old
outstanding problem of nuclear physics.\\
\indent Another longstanding problem is the continuing discrepancy between
 experimental
results and theoretical predictions for $K^+$-nucleus total cross-sections.
Since the $K^+$ interact very weakly (at the hadronic scale) with
 nucleons\cite{do},
they reach the dense central regions of nuclei and thus should be
sensitive to the in-medium modifications of the nucleons and mesons.\\
\indent Different types of such medium effects have been studied. Historically,
the first one has been the swelling of the nucleons in nuclear
 matter\cite{skg}.
This possible swelling has then been related to a decreasing of the mesons
effective mass in the cloud around the nucleon core. This scaling of the mesons
mass with density leads to improved $K^+$-nucleus cross sections\cite{br,cl}.
Let us mention that
 an improvement has also been obtained, but to a lesser extent,
 taking into account
the decreasing of the nucleon effective mass by the scalar nuclear field in
the target, decreasing which modifies the KN amplitude in the
 medium\cite{cl3}.\\
\indent The second type of medium effects which has been considered is the
 possibility
for the  $K^+$ to scatter on the mesons exchanged between the target
 nucleons\cite{ak,jk}. Although these calculations seem not completely under
control since the exchanged mesons are well off-shell\cite{ha},
 they seem to lead also
to a small improvement\cite{jk}.\\
\indent About these medium effects, the most important arises from the
modification of the properties of the mesons exchanged between the $K^+$
and the target nucleons. Since they are well off-mass-shell, it is not
sufficient to consider that the only modification in their propagation is
the change of their effective mass. For example, since they are space-like,
they can produce particle-hole excitations, and thus acquire also a width.
To go beyond this effective mass approximation, we can consider that, at
the hadronic level, this modification of the mesons propagation in nuclear
matter arises from the coupling of these mesons with particle-hole and
nucleon-antinucleon excitations. Since the interaction is strong, this
coupling must be summed up to all orders and thus the choice of the NN
interaction which will be used is very important if we want the final
result might have a chance to be realistic. The question arises if the
relativistic Brueckner-Hartree-Fock interaction we spoke about, which
reproduces the properties of nuclear matter and finite nuclei, could
also give a good description of the polarization of nuclear matter
(at least in the region of small energy-momentum transfer) and thus
lead to improved $K^+$-nucleus cross-sections.\\
\indent In this work, we have calculated the $K^+$-nucleus cross-sections
 using
a KN amplitude in which the mesons exchanged are coupled to the polarization
of the medium in the RPA approximation. This polarization has been obtained
using the relativistic Brueckner-Hartree-Fock NN interaction of Brockmann
and Machleidt\cite{bm}.\\
\indent We have analyzed these effects on the ratio $R_T$ of $K^+-^{12}C$ to
$K^+-d$ total cross sections which has been measured\cite{e,x,p}
from $400 MeV/c$ to $900 MeV/c$.\\
\begin{equation}
\label{rt}
R_T = \frac{\sigma_{tot}(K^+-^{12}C)}{6 \cdot \sigma_{tot}(K^+-d)}
\end{equation}
As emphasized by many authors, this ratio is less
sensitive to experimental and theoretical uncertainties than, for example,
differential cross sections, and thus more transparent to the underlying
physics. \\
\indent The total $K^+$-nucleus cross section has been obtained from the
forward scattering amplitude using the optical theorem. The optical potential
we used to calculate the  $K^+$-nucleus amplitude has been built by folding
the in-medium KN amplitude by the nuclear density.
 The point-like proton distribution of the $^{12}C$ nucleus required in
the
present analysis is that deduced, after the proton finite-size correction has
been made, from the electron-scattering charge density of Sick and
 McCarthy\cite{20} and we have chosen equal n and p-distributions.
For the $K^+$-nucleon amplitude in free space, we have used here the full
Bonn boson exchange model\cite{bo} which is actually one of the more
 elaborate
descriptions of the KN interaction.
In nuclear matter, this amplitude has been modified in order to
take into account the coupling of the $\sigma$ and $\omega$  mesons,
whose exchange provides the dominant part of the
medium-range KN interaction, to the
polarization of the Fermi and Dirac seas. This polarization, calculated
in the one-loop approximation, has been summed up to all orders in the
mesons propagators (RPA approximation) using the NN interaction of
Brockmann and Machleidt.
 Details of the method will appear
 elsewhere\cite{cl2}.
 The heavier particles exchanged lead to very
short-ranged processes less influenced by the nuclear environment.
A Pauli blocking for the nucleon intermediate states has been introduced
 in the integral equation generating the in-medium KN amplitude from the
boson-exchange kernel.
 For the calculation of the
$K^+$-deuteron cross section, since the densities involved are small\cite{21}
 and since
in a nucleus made up of two nucleons it is not possible to excite more than
one particle-hole pair, we have used the free-space KN interaction.
To be consistent, we have thus also used the free-space KN interaction
in the $K^+$-$^{12}C$ calculation for densities lower than half the
saturation density and we have verified that a variation of this cutoff
around this value doesn't change significantly the results.\\
\indent The ratio $R_T$ obtained when the polarization of the nuclear medium
 is taken into account as indicated above, is shown fig.1, curve b. We can
see that this effect is important since the $R_T$ ratio is now, in average,
$\sim$5\%
higher than that calculated using the free-space KN interaction (curve a) and
that the agreement with experiment is considerably improved over the full
energy range. We have used here the potential B of ref.\cite{bm} for the
NN interaction but potentials A and C lead to very similar results.\\
\indent Therefore, it seems that the relativistic
 Brueckner-Hartree-Fock
NN interaction obtained by Brockmann and Machleidt from a free-space
boson exchange model, which gives a good description of nuclear matter
and of finite nuclei, leads also to a polarization of nuclear matter
good enough to reproduce realistically the dressing of the
$\sigma$ and $\omega$ mesons,
at least in the small energy-momentum space-like region which is the
dominant one for forward projectile-nucleus scattering.\\
\indent To our knowledge, this is the first time that such an agreement
for the $R_T$ ratio
is obtained from a fully microscopic (at the hadronic level) calculation.
We believe that this result, obtained without any free parameter,
clearly shows that the main part of the
physical content of the $K^+$-nucleus interaction in the forward
direction has been understood.

\newpage

\begin{center}
{\bf Figure captions}
\end{center}
\bigskip

Fig.1: Ratio $R_T$ of the $K^+-^{12}C$ and $K^+-d$ total cross sections
as a function of $p_{lab}$ calculated, curve (a): with the free-space $K^+N$
interaction, curve (b): with a density-dependent $K^+N$ interaction taking
into account the coupling of the $\sigma$ and $\omega$  mesons with
the polarization of nuclear matter calculated in the RPA approximation.
  The experimental
points are taken from ref.\cite{e} (squares), from ref.\cite{x} (circles) and
from ref.\cite{p} (triangles).

\newpage
 \begin{thebibliography}{99}
\bibitem{bm}
 R. Brockmann and R. Machleidt, Phys. Rev. {\bf C42}, 1965 (1990)
\bibitem{bt}
 R. Brockmann and H. Toki, Phys. Rev. Lett. {\bf 68}, 3408 (1992)
\bibitem{do}
C.B. Dover and G.E. Walker, Phys. Reports {\bf 89},1 (1982) and
references therein
\bibitem{skg}
P. B. Siegel, W. B. Kaufmann and W. R. Gibbs, Phys. Rev. {\bf C31},2184 (1985)
\bibitem{br}
G. E. Brown, C. B. Dover, P. B. Siegel and W. Weise, Phys. Rev.
Lett. {\bf 60},2723 (1988)
\bibitem{cl}
J. C. Caillon and J. Labarsouque, Phys. Lett. {\bf B295}, 21 (1992)
\bibitem{cl3}
J. C. Caillon and J. Labarsouque, Phys. Rev. {\bf C45}, 2503 (1992)
\bibitem{ak}
S. V. Akulinichev, Phys. Rev. Lett. {\bf 68}, 290 (1992)
\bibitem{jk}
M. F. Jiang and D. S. Koltun, Phys. Rev. {\bf C46}, 2462 (1992)
\bibitem{ha}
D. R. Harrington, to be published
\bibitem{e}
D. Bugg et al., Phys. Rev. {\bf 168},1466 (1968)
\bibitem{x}
Y. Mardor et al., Phys. Rev. Lett. {\bf 65}, 2110 (1990)
\bibitem{p}
R. A. Krauss et al., Phys. Rev. {\bf C46}, 655 (1992)
\bibitem{20}
I. Sick and J. S. McCarthy, Nucl. Phys. {\bf A150}, 631 (1970)
\bibitem{bo}
R. Bttgen et al., Nucl. Phys. {\bf A506}, 586 (1990)
\bibitem{cl2}
J. C. Caillon and J. Labarsouque, to be published
\bibitem{21}
L. Hulthn, Rev. Mod. Phys. {\bf 23},1 (1951)
\end{thebibliography}
\end{document}